\documentclass[aps,prl,twocolumn,showpacs,groupedaddress,superscriptaddress,footinbib,longbibliography]{revtex4-1}

\usepackage{natbib,hyperref}
\hypersetup{colorlinks=true, citecolor=blue, urlcolor=blue, linkcolor=blue}
\usepackage[english]{babel}
\usepackage[utf8]{inputenc}
\usepackage{graphicx}
\usepackage[caption=false]{subfig}
\usepackage{amsmath}
\usepackage{amsfonts}
\usepackage{amssymb}
\usepackage{color,soul}
\setulcolor{red}

\usepackage{xcolor}

\begin{document}


\title{Controlling transport of underdamped particles in two-dimensional driven Bravais lattices}
\author{Aritra K. Mukhopadhyay}
\email{Aritra.Mukhopadhyay@physnet.uni-hamburg.de}
\affiliation{Zentrum f\"ur Optische Quantentechnologien, Fachbereich Physik, Universit\"at Hamburg, Luruper Chaussee 149, 22761 Hamburg, Germany}
\author{Peter Schmelcher}
\email{Peter.Schmelcher@physnet.uni-hamburg.de}
\affiliation{Zentrum f\"ur Optische Quantentechnologien, Fachbereich Physik, Universit\"at Hamburg, Luruper Chaussee 149, 22761 Hamburg, Germany}
\affiliation{The Hamburg Centre for Ultrafast Imaging, Universit\"at Hamburg, Luruper Chaussee 149, 22761 Hamburg, Germany}

\date{\today}

\begin{abstract}
We demonstrate the directed transport of underdamped particles in two dimensional lattices of arbitrary geometry driven by an unbiased ac-driving force. The direction of transport can be controlled via the lattice geometry as well as the strength and orientation of the oscillating drive. The breaking of the spatial inversion symmetry, which is necessary for the emergence of directed transport, is achieved solely due to the structure and geometry of the lattice. The most important criterion determining the transport direction is shown to be the ballistic attractors underlying the phase space of our weakly dissipative non-linear dynamical system. This allows the prediction of transport direction even for setups like driven oblique lattices where the standard symmetry arguments of transport control fail. Our results can be experimentally realized using holographic optical lattice based setups with colloids or cold atoms.
\end{abstract}


\maketitle

\paragraph*{Introduction.\textemdash}The interplay between nonlinearity and symmetry breaking in an unbiased non-equilibrium environment has been shown to rectify random particle motion into unidirectional particle transport - a phenomenon usually referred to as the `ratchet effect' \cite{Astumian1997,Julicher1997,Flach2000,Denisov2014,Prost1994,Magnasco1993,Bartussek1994,Faucheux1995}. It was initially conceived as a working principle to describe the performance of various biological motors \cite{Astumian2002,Ait-Haddou2003}. However, today the ratchet effect has attracted widespread interests and has found applications across various disciplines like biological, atomic and condensed matter physics \cite{Hanggi2005,Renzoni2009,Denisov2014,Cubero2016a,Reichhardt2017}. Different schemes based on this mechanism have been implemented to control, among others, the topological soliton dynamics in ionic crystals \cite{Brox2017}, design electron transport in organic semiconductors \cite{Roeling2011} and organic bulk heterojunctions \cite{Kedem2017}, control diffusion of driven magnetic nanoparticle \cite{Stoop2019}, realize unidirectional motion of active matter \cite{Ai2016,Reichhardt2017}, rectify voltage in superconducting quantum interference devices (SQUID) \cite{Zapata1996,Spiechowicz2015} and induce transport of fluxons in Josephson junction arrays \cite{Falo1999,Zolotaryuk2012} or vortices in conformal crystal arrays \cite{Reichhardt2016,Reichhardt2015}.

Due to such a widespread applicability of ratchet based transport, unsurprisingly a vast body of literature has been devoted to control the strength and direction of the ratchet current. While the ratchet setups in one spatial dimensional (1D) address only forward or backward transport of particles \cite{Reimann2002,Hanggi2009,Flach2000,Schanz2001,Schanz2005,Denisov2014,Wulf2012,Liebchen2012,Mukhopadhyay2016}, two dimensional (2D) setups allow for transport at arbitrary angles. It has been shown that particles driven via external time periodic forces on a spatially periodic 2D lattice not only allows directed transport parallel to the drive but also at an angle relative to the driving law \cite{Arzola2017,Mukhopadhyay2018a} or even completely orthogonal to it \cite{Reichhardt2003}. Although there has been major technical advancement in experimental realization of such 2D ratchets involving different systems like cold atoms and colloids, there are certain common drawbacks in most of them. Firstly, most of these ratchet based setups in 2D operate in the overdamped regime where the inertial effects can be neglected. However, there exists a large class of systems which do not operate in this overdamped regime, such as self-propelled vibrated particles \cite{Scholz2018}, underdamped colloids \cite{Sancho2015}, gold and polystyrene nanoparticles in optical systems \cite{Shi2018} or granular particles. Although the control of directed transport would be certainly desirable in these systems, an understanding of the ratchet phenomenon in such underdamped 2D setups is lacking. Secondly, a majority of these setups usually require an external static force as a bias in order to realize directed transport of particles. There are very few setups in 2D where the transport is achieved solely due to an unbiased ac-driving force \cite{Mukhopadhyay2018a,Arzola2017}. Finally, almost all of these setups have focused on directed transport in driven square lattices. Only recently it has been shown that lattices with other geometries, especially oblique lattices, also allow directed transport although in the overdamped regime \cite{Arzola2017}.

In this work, we address the above three key limitations of the traditional 2D ratchet setups. Specifically, we show that it is possible to realize directed transport of \textit{underdamped} particles along \textit{designated directions} by externally driving 2D Bravais lattices of different geometries with an \textit{unbiased} time dependent driving force. The necessary breaking of the spatial inversion symmetry in our setup is achieved solely due to the lattice geometry. Any residual reflection symmetry can be optionally broken by a suitable orientation of the driving force. We show that the resulting direction of transport can be controlled and explained in terms of the ballistic attractors underlying the phase space of our dissipative non-linear dynamical system. It is important to stress that generally such a setup does not allow the prediction of the transport direction a priori due to the absence of any line of reflection symmetry. However, we show that it is possible to realize directed transport of particles along specific directions irrespective of the lattice geometry and orientation of the oscillating drive.

\paragraph{Setup.\textemdash} 
We consider $N$ non-interacting classical particles in a 2D dissipative potential landscape $V(x,y)$=$\sum_{m,n=-\infty}^{+\infty} V_{mn} e^{-\beta\left( {\bf r} - {\bf r}_{mn}\right)^2}$ formed by a lattice of 2D Gaussian barriers centered at positions ${\bf r}_{mn}=(mL,nL)$, $m,n$ $\in\mathbb{Z}$, having site-dependent potential heights $V_{mn}$. The lattice is driven by an external harmonic driving force $\mathbf{f}(t)=a\cos \omega t (\cos \theta_d , \sin \theta_d)$. Here, $a$ and $\omega$ are the amplitude and the frequency of the driving respectively and $\theta_d$ denotes the angle of the driving force with respect to the $x$-axis. Introducing dimensionless variables $x'=\frac{x}{L}$, $y'=\frac{y}{L}$ and $t'=\omega t$ and dropping the primes for simplicity, the equation of motion for a single particle at position  ${\bf r}=(x,y)$ with velocity ${\bf \dot{r}}=(\dot{x},\dot{y})$ reads
\begin{eqnarray}
\ddot{\bf r} &=& -\gamma \dot{\bf r} + {\bf F}(t)+ \pmb{\xi} (t)\nonumber \\
 &+&  \sum_{m,n=-\infty}^{+\infty} U_{mn}\left( {\bf r} - {\bf R}_{mn} \right) e^{-\alpha({\bf r} - {\bf R}_{mn})^2} \label{eqm1}
\end{eqnarray}
where ${\bf F}(t)=d\cos t\left(\cos\theta_d, \sin\theta_d\right)$ is the effective site dependent driving law, ${\bf R}_{mn}=(m,n)$ denotes the positions of the maxima of the Gaussian barriers. The different scaled parameters governing the system are: $U_{mn}=\frac{2V_{mn}\beta}{m\omega ^2}$ denoting the effective barrier heights, an effective driving amplitude $d=\frac{a}{m\omega ^2 L}$, an effective dissipation coefficient $\gamma=\frac{\tilde{\gamma}}{m\omega}$ and the parameter $\alpha=\beta L^2$. $\pmb{\xi} (t)=(\xi_x,\xi_y)$ denotes thermal fluctuations modeled by Gaussian white noise of zero mean with the property $\langle \xi_i (t) \xi_j (t')\rangle = 2 D\delta_{ij}\delta (t-t')$ where $i,j \in {x,y}$ and $D=\frac{\tilde{\gamma} k_B \mathcal{T}}{m\omega ^2 L^2}$ is the dimensionless noise strength with $\mathcal{T}$ and $k_B$ denoting the temperature and Boltzmann constant respectively. Our setup is a driven superlattice formed by the superposition of different sublattices each consisting of barriers possessing distinct heights $U_{mn}$. Each sublattice is individually symmetric with respect to the spatial inversion and time shift symmetry $S_{\mathbf{r}}$: $\mathbf{r}\longrightarrow -\mathbf{r} + \pmb{\delta}$, $t\longrightarrow t + \tau$ (for arbitrary constant translations $\pmb{\delta}$ and $\tau$ of space and time respectively). However, the necessary breaking of $S_{\mathbf{r}}$ symmetry required for directed transport is achieved by a superposition of at least three sublattices consisting of barriers with different heights. Since our setup is dissipative, the time reversal and spatial shift symmetry $S_{t}$: $t\longrightarrow -t + \tau$, $\mathbf{r}\longrightarrow \mathbf{r} + \pmb{\delta}$ is also broken.

The setup can be experimentally realized, e.g. by using monodisperse colloidal particles in a 2D lattice obtained by reflecting a linearly polarized laser beam onto a spatial light modulator (SLM) displaying a computer generated hologram which can then be driven using a piezo-modulator \cite{Arzola2017}. A second highly controllable setup could be driven lattices based on holographic trapping of atoms \cite{Nogrette2014,Kim2016,Barredo2016,Stuart2018} in the regime of microkelvin temperatures where a classical description of cold atom ratchets is appropriate \cite{Renzoni2009}.

\begin{figure} 
	\centerline{\includegraphics[width=0.67\textwidth]{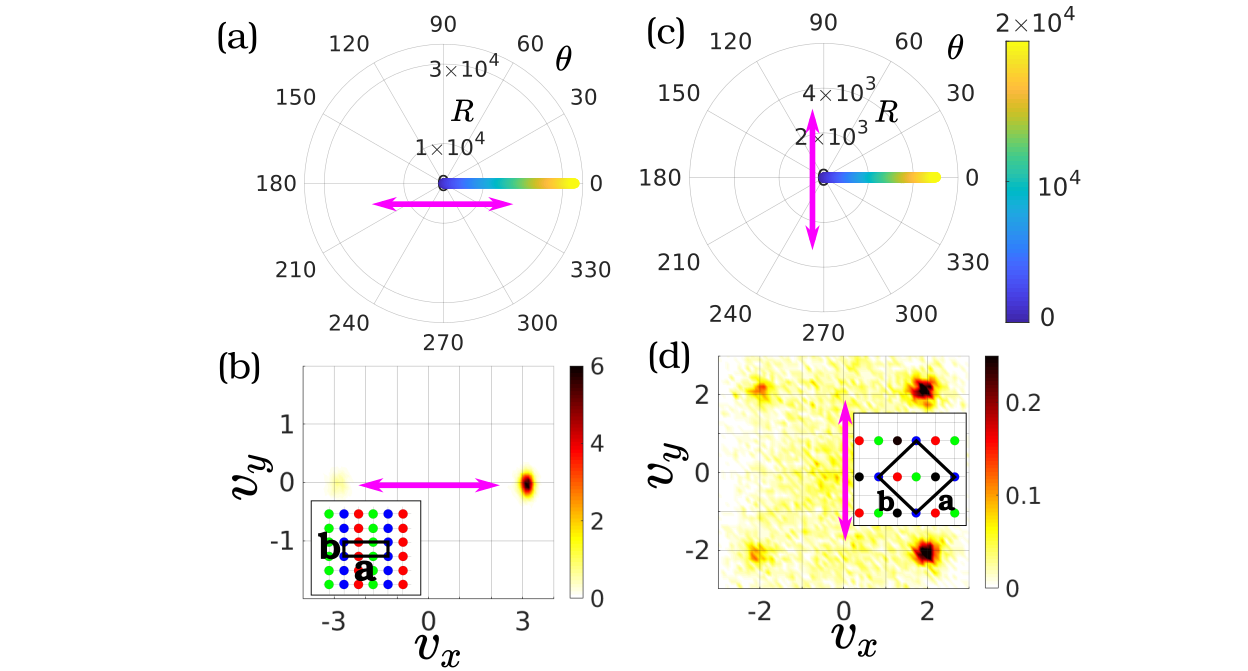}}
	\caption{Mean position of the particle ensemble (in radial $R$ and angular $\theta$ coordinates) as a function of time $t$ (in colorbar) for the (a) rectangular lattice driven along $\theta_d=0^\circ$ with $d=0.3$ and (c) square lattice driven along $\theta_d=90^\circ$ with $d=0.5$. Figures (b) and (d) provide snapshots of the ensemble velocity distribution at $t=t_f$ for the two setups (a) and (c) respectively. The insets in (b) and (d) depict schematic representations of the two corresponding lattices along with the lattice vectors $\mathbf{a}$ and $\mathbf{b}$, with each colored circle denoting the position of an individual Gaussian barrier. The different colors denote different barrier heights (Fig. (b), inset) $U_{mn}=$0.5(blue), 1.0(red), 1.5(green), (Fig. (d), inset) $U_{mn}=$0.3(blue), 0.6(red), 0.9(green), 1.2(black). The driving axis is denoted by the pink double arrowed lines. Remaining parameters: $\gamma=10^{-2}$, $\alpha=5$ and $D=1.5\times 10^{-4}$. }\label{fig1}
\end{figure}

In order to explore the particle transport in our lattice characterized by the average velocity of the particle ensemble, we initialize $N=10^4$ particles within a square region $x,y \in [-10,10]$ with small random velocities $v_x,v_y \in [-0.1,0.1]$, such that their initial kinetic energies are small compared to the potential height of the Gaussian barriers. Subsequently we time evolve our ensemble up to time $t_f= 2\times 10^4$ by numerical integration of Eq.~\ref{eqm1}. For all our setups, we consider noise strength $D>0$ and the resulting asymptotic transport velocity is independent of the specific initial conditions of each particles. We demonstrate that it is possible to realize particle transport parallel to the driving force axis (axial transport), orthogonal to it (lateral transport) or even in an oblique direction for different lattice geometries using our setup.

\paragraph{Axial transport in a rectangular lattice.\textemdash} 
In our first setup (Figs.~\ref{fig1}a,b), we consider a rectangular superlattice constructed by superposing three rectangular lattices with lattice vectors $\mathbf{a}=(3,0)$ and $\mathbf{b}=(0,1)$. Although the setup breaks inversion symmetry $S_{\mathbf{r}}$, it is invariant under $P_y: y\longrightarrow -y$ rendering any line parallel to the $x$-axis as the line of reflection symmetry modulo a spatial translation along the $y$ direction. The lattice is driven along this symmetry axis by choosing $\theta_d=0^\circ$. As expected from the symmetry argument, the particles exhibit no transport along the $y$-direction and a net directed transport is observed along the positive $x$-direction (Fig.~\ref{fig1}a). The direction of transport can be better understood by analyzing the asymptotic velocity distribution of the particles at the end of the simulation time i.e. $t_f$, which shows that most particles travel along the $x$-axis with velocities either $\mathbf {v} \approx (3,0)$ or $(-3,0)$ (Fig.~\ref{fig1}b). For the chosen parameter regime, these velocities correspond to the average velocities of the two ballistic attractors denoting synchronized motion of particles through the oscillating lattice traveling one unit cell per unit time either parallel or anti-parallel to the lattice vector $\mathbf{a}$ in the deterministic limit $D\rightarrow 0$. Even for $D>0$, the attractors are not completely destroyed and at longer timescales the particles move approximately with the same velocities as the average velocities of these attractors. However due to the broken $P_x: x\longrightarrow -x$ symmetry, the velocity distribution is asymmetric and significantly more particles travel towards right than towards left resulting in an axial transport along the positive $x$-direction.

\paragraph{Lateral transport in a square lattice.\textemdash} 
Lattices possessing a line of reflection symmetry can also exhibit directed transport along a direction orthogonal to the driving force. To illustrate this, we consider a square lattice formed by the superposition of four square lattices with lattice vectors $\mathbf{a}=(2,2)$ and $\mathbf{b}=(-2,2)$ (Fig.~\ref{fig1}c,d). This lattice too breaks both $S_{\mathbf{r}}$ and $P_x$ symmetries but preserves the $P_y$ symmetry. Upon driving the lattice along the $y$-axis, which is orthogonal to the symmetry axis, a lateral current is observed along the positive $x$ direction in accordance with the symmetry argument (Fig.~\ref{fig1}c). From the peaks of the asymptotic velocity distribution of the particles (Fig.~\ref{fig1}d), it is evident that the underlying particle dynamics is governed mainly by the four ballistic attractors with average velocities $(2,2)$, $(2,-2)$, $(-2,2)$ and $(-2,-2)$. These correspond to particles exhibiting regular motion, moving one unit cell per unit time along directions parallel and anti-parallel to the two lattice vectors $\mathbf{a}$ and $\mathbf{b}$. The $P_y$ symmetry is clearly reflected in the asymptotic velocity distribution due to which almost equal number of particles possess $v_y>0$ and  $v_y<0$, thus prohibiting any average transport in the $y$ direction (Fig.~\ref{fig1}d). However due to the $P_x$ symmetry breaking, the number of particles moving along the positive $x$ direction is much higher and hence directed transport occurs along this direction.

\begin{figure}
	\centerline{\includegraphics[width=0.67\textwidth]{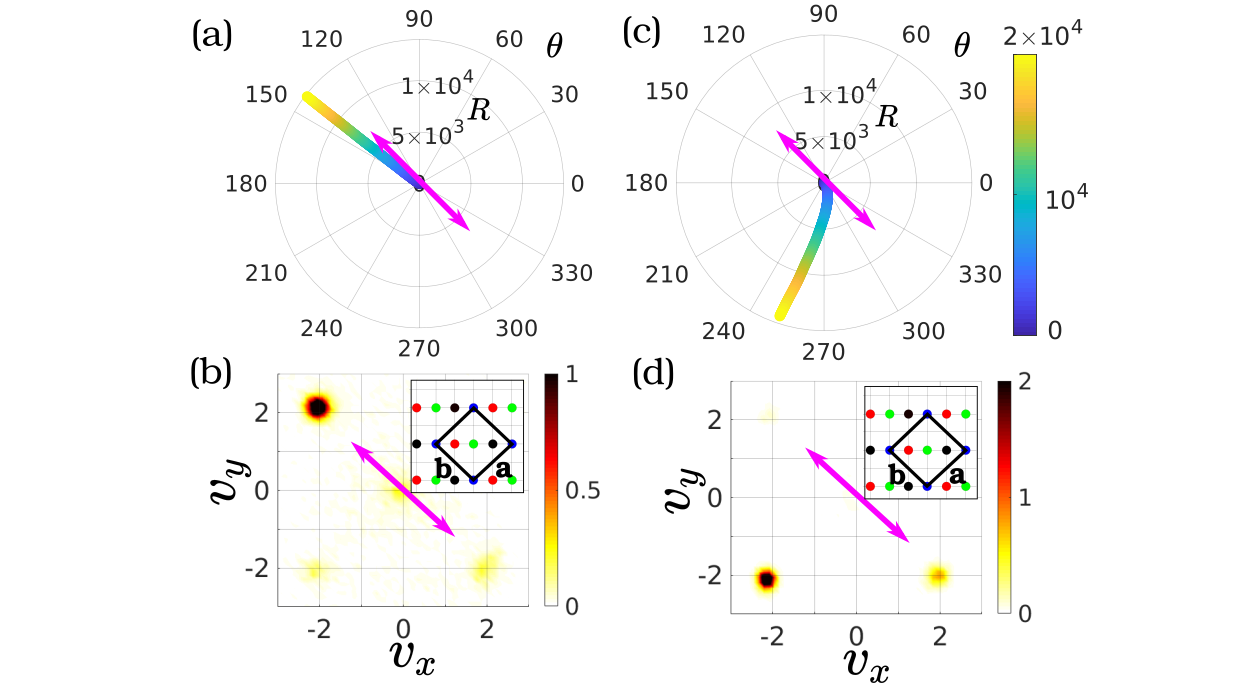}}
	\caption{Mean position of the particle ensemble (in radial $R$ and angular $\theta$ coordinates) as a function of time $t$ (in colorbar) for the same square lattice setup as in Fig. 1(c,d) driven along $\theta_d = 135^\circ$ (pink double arrowed lines) with (a) $d=0.6$ and (c) $d=1.0$. Figures (b) and (d) provide snapshots of the ensemble velocity distribution at $t=t_f$ for the two setups (a) and (c) respectively; and the insets depict schematic representations of the corresponding lattice. Remaining parameters are the same as in Fig. 1 (c,d).}\label{fig2}
\end{figure}

\paragraph{Driving induced breaking of the reflection symmetry.\textemdash}

The residual reflection symmetry $P_y$ in our rectangular or square lattice setups can be broken by driving the lattice oblique to the line of reflection symmetry. Since $P_y$ transforms $\theta_d \rightarrow -\theta_d$, hence ${\bf F}(t) \rightarrow  {\bf \tilde{F}}(t)=d\cos t\left(\cos\theta_d, -\sin\theta_d\right)$ which cannot be transformed back to ${\bf F}(t)$ by any additional time shift operation for $\theta_d \neq 0^\circ,90^\circ,180^\circ$ or $270^\circ$. To illustrate this, we consider the same square lattice as in Fig.~\ref{fig1}(c,d) but now driven along the lattice vector $\mathbf{b}$ by choosing $\theta_d = 135^\circ$ (Fig.~\ref{fig2}). Although the broken $S_{\mathbf{r}}$ symmetry ensures the existence of directed transport, the direction of transport can no longer be predicted from symmetry considerations alone. However, we show that it is possible to control the underlying ballistic attractors and hence the transport direction by varying the amplitude of the driving force $d$. For $d=0.6$, the ensemble is transported along $\theta \approx 140^\circ$ almost parallel to the driving force along the lattice vector $\mathbf{b}$ (Fig.~\ref{fig2}a). The peak at $\mathbf{v} \approx (-2,2)$ in the asymptotic particle velocity distribution shows that the asymptotic dynamics of the ensemble is governed primarily by a single ballistic attractor with average velocity $(-2,2)$ (Fig.~\ref{fig2}b) denoting synchronized particle motion parallel to $\mathbf{b}$. Therefore directed transport appears along this direction. Upon driving the lattice along the same axis, but with a higher driving amplitude $d=1.0$, the direction of transport can be rotated to an almost perpendicular direction $\theta \approx 250^\circ$ (Fig.~\ref{fig2}c). The change in the driving strength changes the dominant attractor governing the transport, which now has an average velocity $(-2,-2)$, propelling majority of the particles to move with this velocity in a direction anti-parallel to the lattice vector $\mathbf{a}$ (Fig.~\ref{fig2}d), hence explaining the transport. We note that for a broad range of value of $d$, the particle dynamics is governed by the four ballistic attractors with average velocities $(2,2)$, $(2,-2)$, $(-2,2)$ and $(-2,-2)$ (see Supplemental Material). The transport direction is determined by the attractor with the highest asymptotic particle occupancy. Hence for different values of $d$, directed transport occurs to a good approximation along one of these four directions.

\begin{figure}
	\centerline{\includegraphics[width=0.67\textwidth]{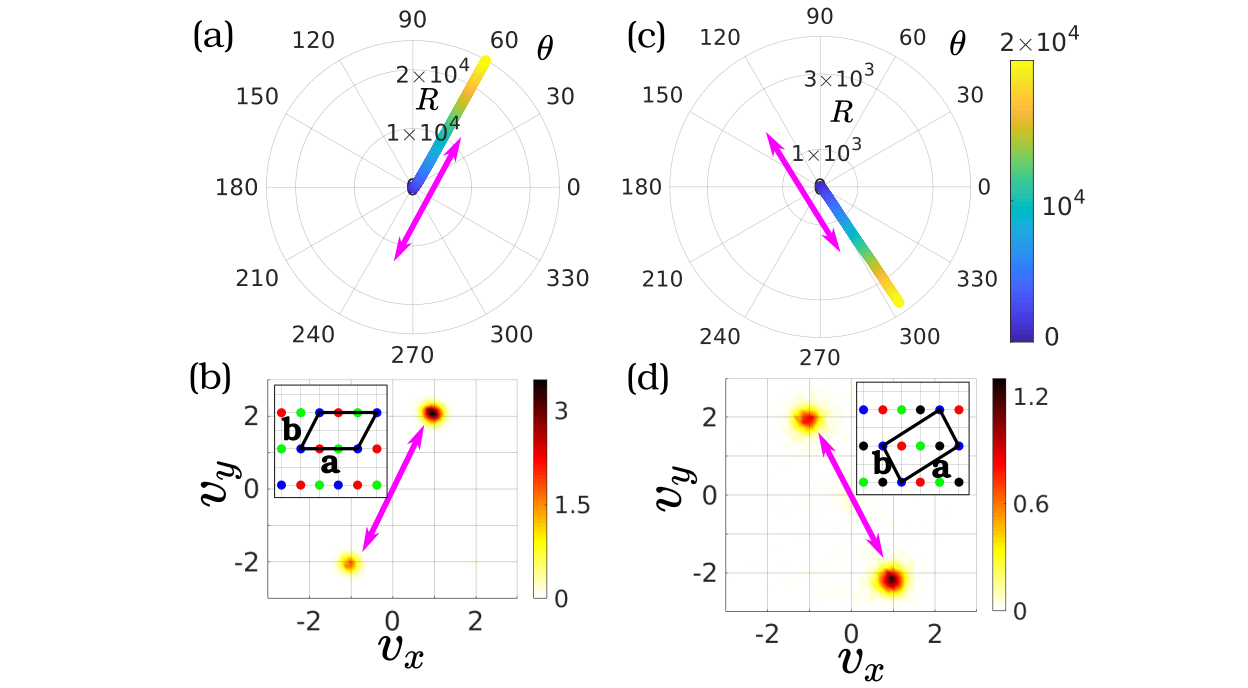}}
	\caption{Mean position of the particle ensemble (in radial $R$ and angular $\theta$ coordinates) as a function of time $t$ (in colorbar) for the two oblique lattices (a) \textit{O1} driven along $\theta_d\approx 63^\circ$ with $d=0.75$ and (c) \textit{O2} driven along $\theta_d\approx 116^\circ$ with $d=0.6$. Figures (b) and (d) provide snapshots of the ensemble velocity distribution at $t=t_f$ for the two setups (a) and (c) respectively. The insets in (b) and (d) depict schematic representations of the two corresponding lattices with different colored circles denoting different barrier heights:  $U_{mn}=$0.3(blue), 0.6(red), 0.9(green), 1.2(black). The driving axis is denoted by the pink double arrowed lines. Remaining parameters: $\gamma=10^{-2}$, $\alpha=5$ and $D=2\times 10^{-4}$.}\label{fig3}
\end{figure}

\paragraph{Oblique lattice.\textemdash} 
In contrast to square and rectangular lattices, a 2D oblique lattice does not possess any lines of reflection symmetry and therefore has no obvious symmetry direction along which directed transport should occur. Even for such a setup we can realize directed transport of particles along a particular direction, specifically along the shortest lattice vector, by controlling the underlying ballistic attractors determining the transport. We illustrate this by considering an oblique lattice \textit{O1} composed of three superimposed oblique lattices with lattice vectors $\mathbf{a}=(3,0)$ and $\mathbf{b}=(1,2)$ ($|\mathbf{a}|>|\mathbf{b}|$) with an angle of approximately $63^\circ$ between them. Upon driving the lattice along $\mathbf{b}$, an axial directed transport of particles is observed along $\theta \approx 63^\circ$ parallel to $\mathbf{b}$ (Fig.~\ref{fig3}a). Most of the particles move asymptotically with $\mathbf{v} \approx (1,2)$ or $(-1,-2)$, which denote the average velocities of the ballistic attractors corresponding to particles moving one unit cell per unit time parallel or anti-parallel to the shortest lattice vector $\mathbf{b}$ (Fig.~\ref{fig3}b). The spatial asymmetry due to the breaking of $S_{\mathbf{r}}$ symmetry is responsible for a higher number of particles moving parallel to $\mathbf{b}$, resulting in the transport along this direction.

\begin{figure}[t]
	\centerline{\includegraphics[width=0.67\textwidth]{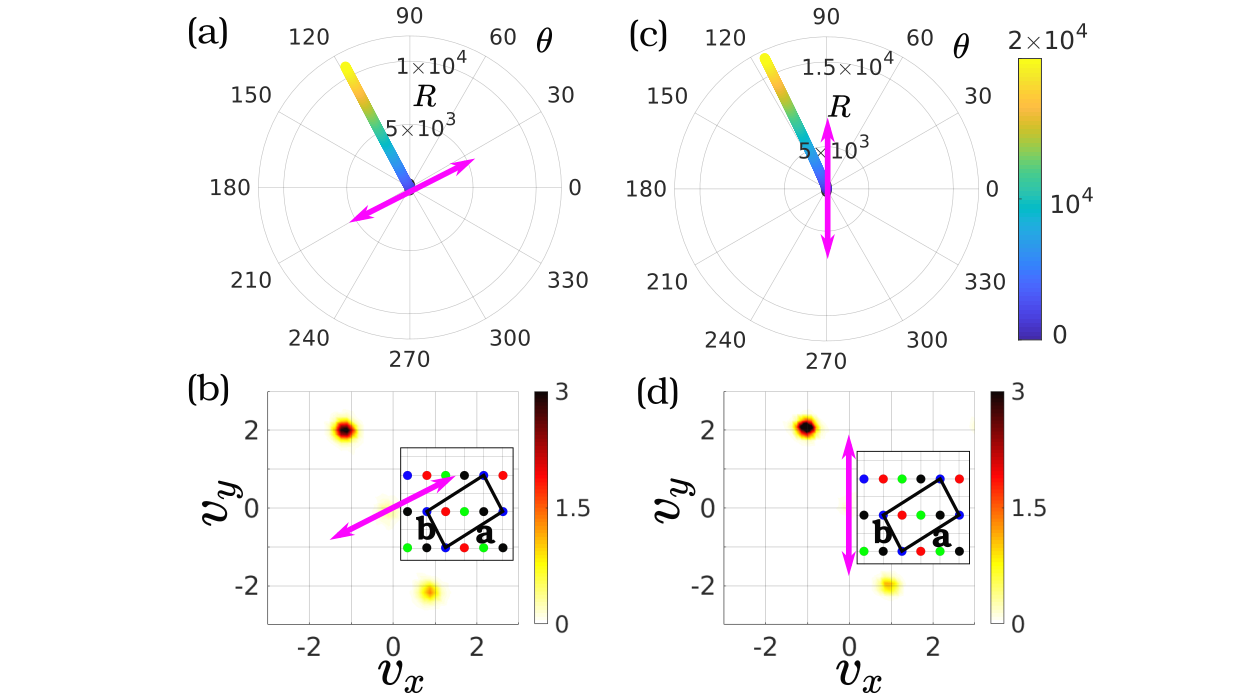}}
	\caption{Mean position of the particle ensemble (in radial $R$ and angular $\theta$ coordinates) as a function of time $t$ (in colorbar) for the oblique lattice \textit{O2} (see Fig. 3(c,d)) driven along (pink double arrowed lines) (a) $\theta_d\approx 26^\circ$ with $d=1.0$ and (c) $\theta_d=90^\circ$ with $d=1.6$. Figures (b) and (d) provide snapshots of the ensemble velocity distribution at $t=t_f$ for the two setups (a) and (c) respectively. The insets depict schematic representations of the corresponding lattices. Remaining parameters are the same as in Fig. 3 (c,d).}\label{fig4}
\end{figure}

Next, we demonstrate that it is possible to direct the particle transport either parallel or anti-parallel to the shortest lattice vector of an oblique lattice irrespective of the direction of the driving force. To illustrate this, we consider a second oblique lattice setup \textit{O2} constructed by the superposition of four oblique lattices having lattice vectors $\mathbf{a}=(3,2)$ and $\mathbf{b}=(-1,2)$ (hence $|\mathbf{a}|>|\mathbf{b}|$) such that the angle between them is approximately $83^\circ$. When the lattice is driven along the lattice vector $\mathbf{b}$ with $d=0.6$, an axial directed transport is observed at $\theta \approx 300^\circ$ anti-parallel to $\mathbf{b}$ (Fig.~\ref{fig3}c). However, upon driving the lattice along an axis perpendicular to $\mathbf{b}$ with $d=1$, a reversal of transport occurs and the ensemble moves along $\theta \approx 120^\circ$ almost parallel to $\mathbf{b}$ thereby exhibiting lateral transport (Fig.~\ref{fig4}a). A directed transport along $\theta \approx 120^\circ$ is also observed when the external drive is along the $y$-axis and $d=1.6$, thus allowing us to realize oblique transport with respect to the driving force (Fig.~\ref{fig4}c). The transport in all these three scenarios is governed by the two ballistic attractors having average velocities $(-1,2)$ and $(1,-2)$ around which the asymptotic velocity distribution of the particles is localized (Figs.~\ref{fig3}d and ~\ref{fig4}(b,d)). Similar to the setup \textit{O1}, these velocities corresponds to synchronized motion of particles moving either parallel or anti-parallel to the shortest lattice vector $\mathbf{b}$ and the transport direction is determined by the relative asymmetry in the number of particles moving along these two directions. Since the dominant ballistic attractors for both the setups \textit{O1} and \textit{O2} remain the same as those in Figs.~\ref{fig3} and ~\ref{fig4} upon varying $d$, the direction of transport too does not change considerably (see Supplemental Material).

\paragraph{Conclusions.\textemdash} 
We have demonstrated the control of directed transport of underdamped particles along specific directions in different types of 2D Bravais lattices driven by unbiased external forces. Most importantly, we show that it is possible to direct the transport along one of the lattice vectors in setups without any line of reflection symmetry irrespective of the driving axis. These setups preclude any prediction of the transport direction a priori based on the standard symmetry arguments. Yet we show that the direction of transport can be well understood in terms of the attractors controlling the asymptotic dynamics of our non-linear dynamical system. The observed directions of transport persists for noise strengths up to $D\lesssim 10^{-3}$ typical for cold atoms or underdamped colloids. The fact that different lattice geometries can be realized simply by varying the potential heights of the Gaussian barriers constituting the sublattices should also allow for a time-dependent control of the transport direction using dynamic holographic optical tweezers \cite{Curtis2002} or dynamical digital hologram generation techniques \cite{Reicherter2006,Lanigan2012}. Future perspectives include the investigation of the impact of the lattice geometry on the chaotic transport in very weakly dissipative and pure Hamiltonian regime and relevant technological applications like development of miniature devices helpful for colloidal sorting or targeted drug delivery.

\begin{acknowledgments}
 A.K.M acknowledges a doctoral research grant (Funding ID: 57129429) by the Deutscher Akademischer Austauschdienst (DAAD). The authors thank B. Liebchen for insightful discussions.
\end{acknowledgments}

\bibliographystyle{apsrev4-1}
\bibliography{mybib}

\end{document}


\title{Controlling transport of underdamped particles in two-dimensional driven Bravais lattices}
\author{Aritra K. Mukhopadhyay}
\email{Aritra.Mukhopadhyay@physnet.uni-hamburg.de}
\affiliation{Zentrum f\"ur Optische Quantentechnologien, Fachbereich Physik, Universit\"at Hamburg, Luruper Chaussee 149, 22761 Hamburg, Germany}
\author{Peter Schmelcher}
\email{Peter.Schmelcher@physnet.uni-hamburg.de}
\affiliation{Zentrum f\"ur Optische Quantentechnologien, Fachbereich Physik, Universit\"at Hamburg, Luruper Chaussee 149, 22761 Hamburg, Germany}
\affiliation{The Hamburg Centre for Ultrafast Imaging, Universit\"at Hamburg, Luruper Chaussee 149, 22761 Hamburg, Germany}

\date{\today}
\maketitle

\section{Supplemental Material}

\subsection{Role of driving strength in directed transport}
Here we discuss the behaviour of the attractors underlying the phase space of the different lattices described in the main text for different driving amplitudes $d$. As mentioned in the main text, the deterministic $D\rightarrow 0$ dynamics of our driven lattice setup is governed by the asymptotic attractors in the system, which can be either \textit{chaotic} denoting diffusive particle motion through the lattice or \textit{ballistic} representing regular periodic motion of the particles. The attractors are characterized by their average velocities $\bar{\bf v}$, which for the ballistic attractors can be expressed as $\bar{\bf {v}}= \frac{1}{T}\left( \frac{m_a}{n_a} \mathbf{a} + \frac{m_b}{n_b} \mathbf{b} \right)$ with $m_a,n_a,m_b,n_b \in \mathbb{Z}$, $\mathbf{a},\mathbf{b}$ being the two lattice vectors and $T$ is the temporal driving period which for our case is unity. Although all our setups are characterized by both the chaotic and ballistic attractors, we have focused on the directed transport governed solely by the ballistic attractors. In the following, we discuss the ballistic attractors corresponding to each of our setups for different values of the driving strength $d$. For this, we numerically propagate $N=10^4$ particles within a square region $x,y \in [-10,10]$ with small random velocities $v_x,v_y \in [-0.1,0.1]$ for different values of $d$ up to $t=2\times 10^4$. The asymptotic average velocities of each of these trajectories correspond to the average velocity $\bar{\bf {v}}$ of the different attractors underlying the setup. When $\bar{\bf {v}}$ is expressed in polar coordinates, the angular component $\bar{\theta}$ denotes the average direction of an attractor and the modulus $|\bar{\bf {v}}|$ denotes its average speed.

\begin{figure}[b]
	\includegraphics[width=0.6\textwidth]{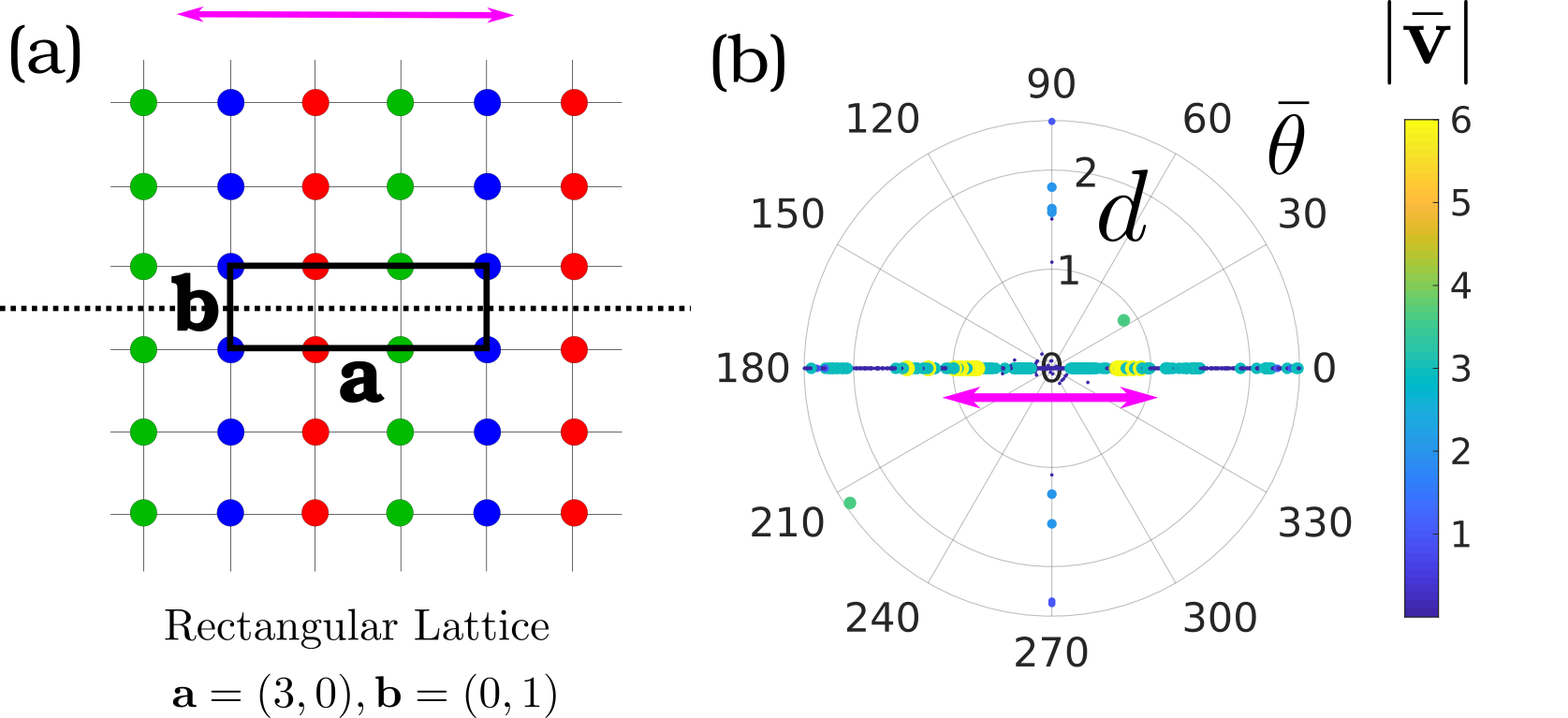}
	\caption{(a) Schematic representation of our rectangular lattice setup with lattice vectors $\mathbf{a}$ and $\mathbf{b}$. Each filled colored circles represent Gaussian barriers with different heights $U_{mn}=$0.5(blue), 1.0(red),1.5(green). The dotted line parallel to the $x$-axis denotes the line of reflection symmetry. (b) The average direction of the attractors $\bar{\theta}$ (in angular coordinates) as a function of driving amplitude $d$ (in radial coordinates) with the colorbar denoting their average speed $|\bar{\bf {v}}|$. The driving axis is denoted by the pink double arrowed lines in both the figures. Remaining parameters: $\gamma=10^{-2}$,$\theta_d=0^\circ$, $\alpha=5$ and $D=1.5\times 10^{-4}$.}\label{fig1}
\end{figure}

\subsubsection{Rectangular lattice:} 
First, we consider the rectangular lattice with spatial period $(3,1)$ and lattice vectors $\mathbf{a}=(3,0)$ and $\mathbf{b}=(0,1)$ driven along the $x$-axis (Fig.~\ref{fig1}a) as described in the main text. For different values of the driving strength $d$, we note that the majority of the attractors are located either along $\bar{\theta}=0^\circ$ or $\bar{\theta}=180^\circ$ (Fig.~\ref{fig1}b) with some isolated ones along $\theta = 30^\circ, 90^\circ, 210^\circ, 270^\circ$. Therefore, for almost any value of $d$, we would expect most of the particles in our setup to move either along the positive or negative $x$ direction. The breaking of the spatial inversion symmetry induces an asymmetry in the number of particles moving in the two directions, therefore directed transport emerges along any one of them as discussed in the main text. The reflection symmetry in the $y$ direction forbids any transport along the $y$-direction.

\begin{figure}
	\includegraphics[width=0.6\textwidth]{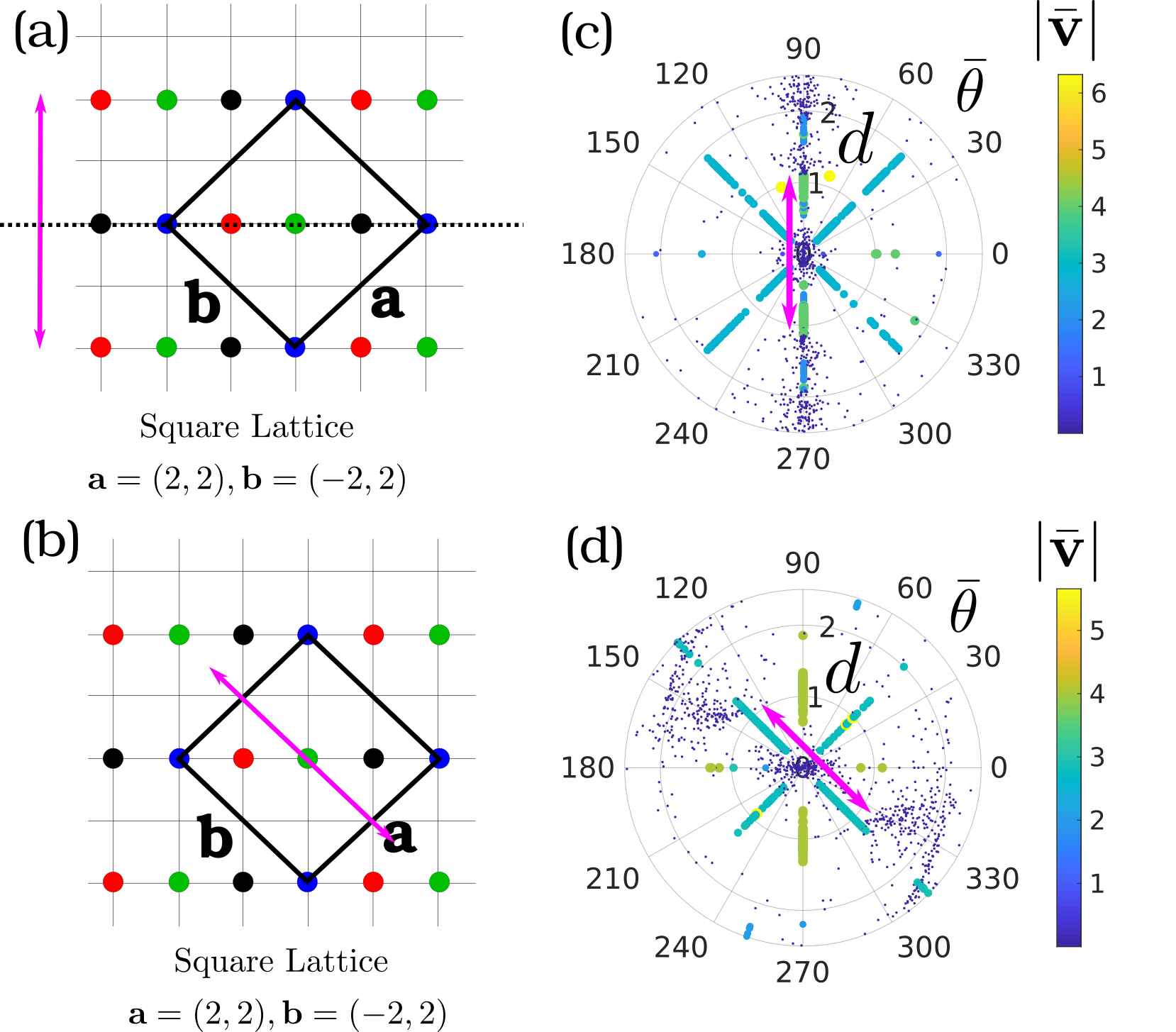}
	\caption{Schematic representation of our square lattice setup with lattice vectors $\mathbf{a}$ and $\mathbf{b}$ driven along (a) $y$-axis with $\theta_d=90^\circ$ and (b) $\theta_d=135^\circ$. Each filled colored circles represent Gaussian barriers with different heights $U_{mn}=$0.3(blue), 0.6(red), 0.9(green), 1.2(black). The dotted line in (a) denotes the line of reflection symmetry. (c) and (d) the average direction of the attractors $\bar{\theta}$ (in angular coordinates) as a function of driving amplitude $d$ (in radial coordinates) with the colorbar denoting their average speed $|\bar{\bf {v}}|$ corresponding to (a) and (b) respectively. The blue colored smaller dots denote the chaotic attractors whereas the larger filled circles of other colors denote the ballistic attractors. The driving axis is denoted by the pink double arrowed lines in all the figures. Remaining parameters: $\gamma=10^{-2}$, $\alpha=5$ and $D=1.5\times 10^{-4}$.}\label{fig2}
\end{figure}

\subsubsection{Square lattice:}
For our square lattice setup with lattice vectors $\mathbf{a}=(2,2)$ and $\mathbf{b}=(-2,2)$ (Fig.~\ref{fig2}(a,b)), there exist roughly six different directions corresponding to the ballistic attractors irrespective of the driving axis and driving strength. These are along $45^\circ$, $90^\circ$, $135^\circ$, $225^\circ$, $270^\circ$ and $315^\circ$ (Fig.~\ref{fig2}(c,d)). However in the presence of noise, the breaking of the spatial inversion symmetry induces asymmetric `jumps' of trajectories between different attractors such that only one or two ballistic attractors govern the particle dynamics asymptotically (see Figs. 1 and 2 in the main text). 

When the lattice is driven along the $y$-axis (Fig.~\ref{fig2}a), the reflection symmetry about the $x$-axis ensures that an equal number of particles asymptotically end up in attractors with $0^\circ < \bar{\theta} <180^\circ$ and $180^\circ < \bar{\theta} <360^\circ$, thus prohibiting transport in the  $y$-direction. But the spatial inversion symmetry ensures an imbalance between the number of particles whose dynamics is governed by the ballistic attractors with $\bar{\theta}=45^\circ, 315^\circ$ and those with $\bar{\theta}=135^\circ, 225^\circ$. This ensures a net transport along the positive or negative $x$-direction as discussed in the main text. 

For any other choice of the driving axis, e.g. as in Fig.~\ref{fig2}b, there exists no line of symmetry and usually only one of the six ballistic attractors controls the asymptotic particle dynamics. Hence, for different values of the driving strength $d$, directed transport is observed along one of these directions. In the main text we have shown two such examples, where the transport is governed by the attractors with $\bar{\theta}=135^\circ$ and $225^\circ$ respectively for the same orientation of the driving axis as in Fig.~\ref{fig2}b but for two different driving strengths $d$.

\subsubsection{Oblique lattice:}
\begin{figure}
	\includegraphics[width=0.6\textwidth]{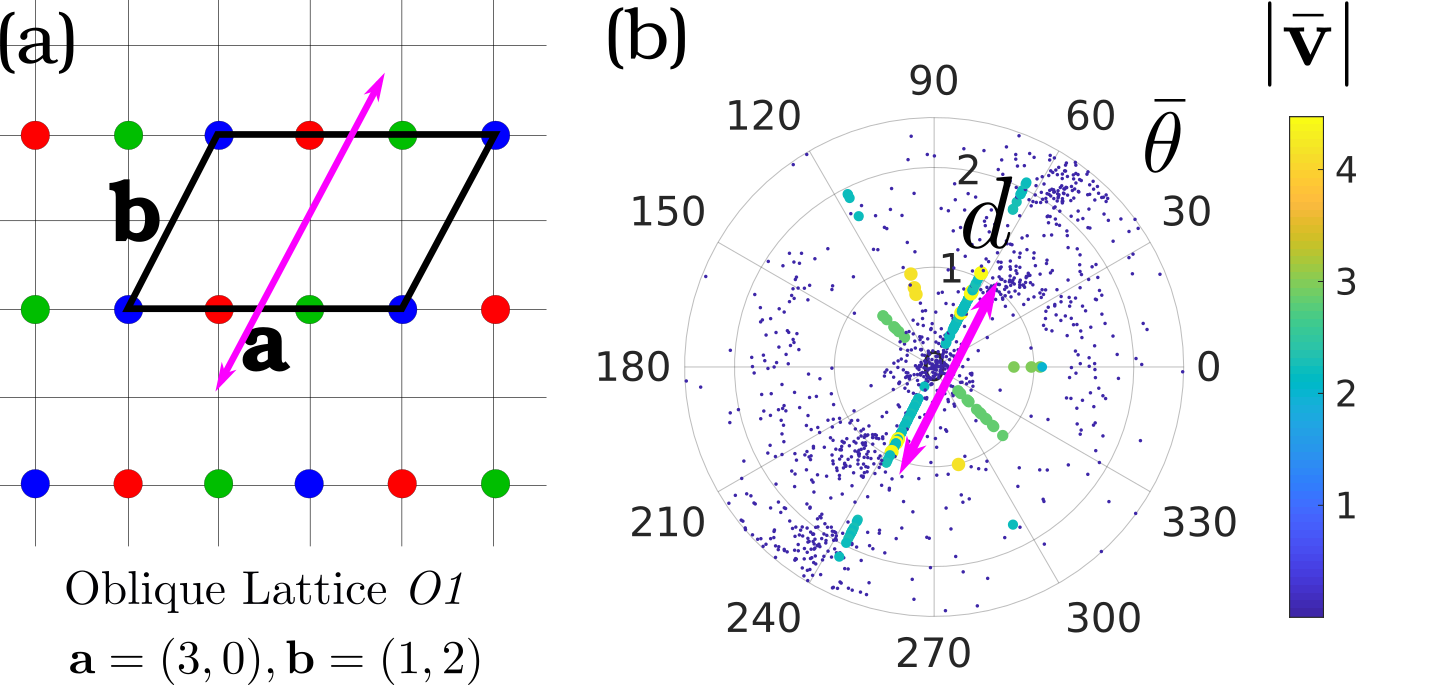}
	\caption{(a) Schematic representation of our oblique lattice setup \textit{O1} with lattice vectors $\mathbf{a}$ and $\mathbf{b}$ driven parallel to $\mathbf{b}$. Each filled colored circles represent Gaussian barriers with different heights $U_{mn}=$0.3(blue), 0.6(red),0.9(green). (b) The average direction of the attractors $\bar{\theta}$ (in angular coordinates) as a function of driving amplitude $d$ (in radial coordinates) with the colorbar denoting their average speed $|\bar{\bf {v}}|$. The blue colored smaller dots denote the chaotic attractors whereas the larger filled circles of other colors denote the ballistic attractors. The driving axis is denoted by the pink double arrowed lines in both the figures. Remaining parameters: $\gamma=10^{-2}$, $\theta_d\approx 63^\circ$, $\alpha=5$ and $D=2\times 10^{-4}$.}\label{fig3}
\end{figure}

\begin{figure}
	\includegraphics[width=0.7\textwidth]{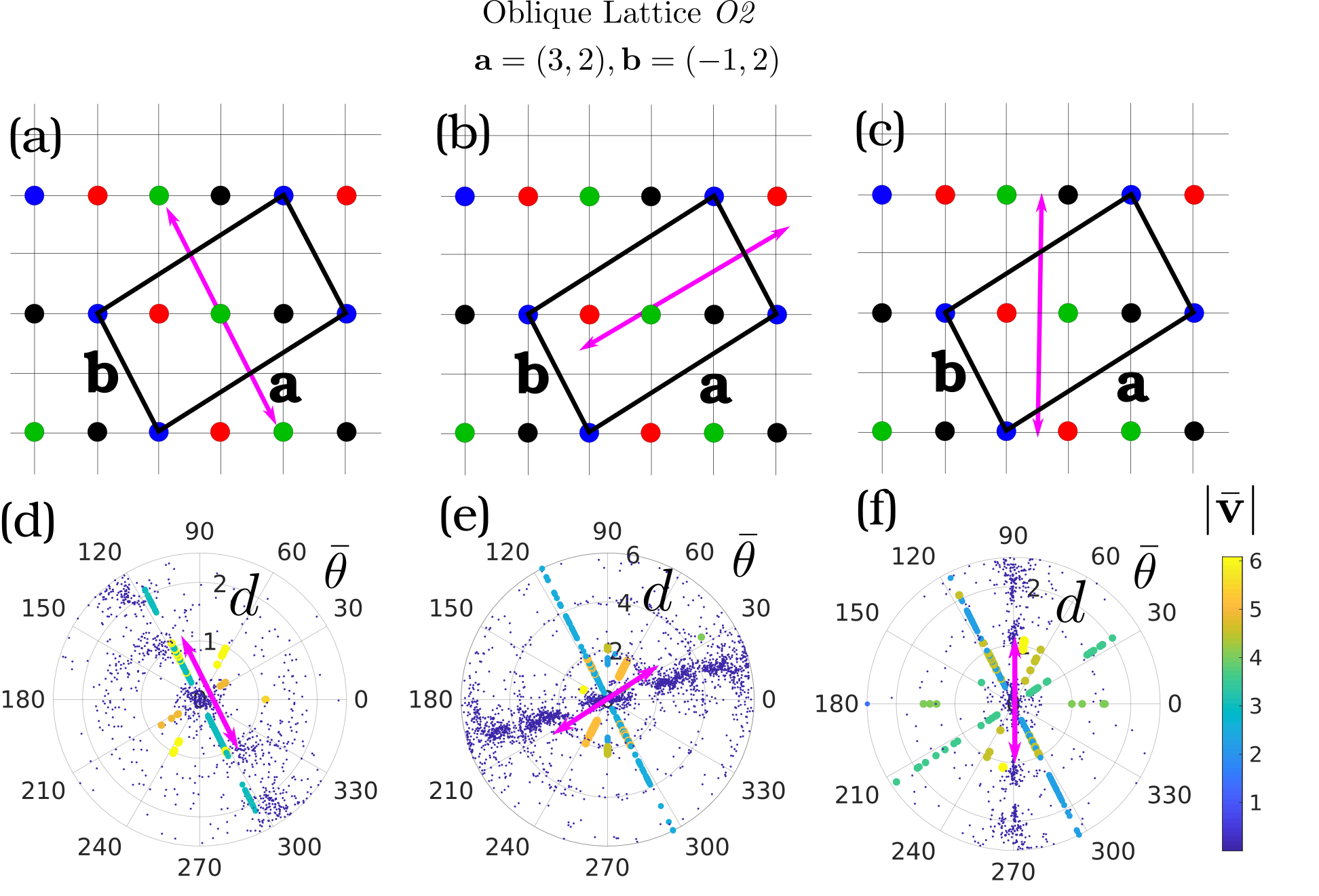}
	\caption{Schematic representation of our oblique lattice setup \textit{O2} with lattice vectors $\mathbf{a}$ and $\mathbf{b}$ driven along (a) $\theta_d \approx 116^\circ$ (b) $\theta_d \approx 26^\circ$ and (c) $\theta_d=90^\circ$. Each filled colored circles represent Gaussian barriers with different heights $U_{mn}=$0.3(blue), 0.6(red), 0.9(green), 1.2(black). (d), (e) and (f) The average direction of the attractors $\bar{\theta}$ (in angular coordinates) as a function of driving amplitude $d$ (in radial coordinates) with the colorbar denoting their average speed $|\bar{\bf {v}}|$ corresponding to (a), (b) and (c) respectively. The blue colored smaller dots denote chaotic attractors whereas the larger filled circles of other colors denote the ballistic attractors. The driving axis is denoted by the pink double arrowed lines in all the figures. Remaining parameters: $\gamma=10^{-2}$, $\alpha=5$ and $D=2\times 10^{-4}$.}\label{fig4}
\end{figure}

For our oblique lattice setups (\textit{O1} and \textit{O2}) with lattice vectors $\mathbf{a}$ and $\mathbf{b}$ ($|\mathbf{a}|>|\mathbf{b}|$), we find that there are always two ballistic attractors oriented parallel and anti-parallel to the smallest lattice vector $\mathbf{b}$ irrespective of the orientation and strength of the driving force (see Figs.~\ref{fig3} and ~\ref{fig4}). As a result for most values of driving strength $d$ and orientation $\theta_d$, it is possible to realize directed transport of particles along any one of these directions as we have discussed in the main text.

Hence for all our setups, the underlying ballistic attractors are quite robust with respect to slight variations of the driving strength $d$. The particular values of $d$ in the main text have been chosen in order to exemplify the directed transport at specific angles in each of these setups.

\bibliography{mybib}